\documentstyle[12pt]{article}
\parskip=\medskipamount

\newcommand {\cD}{{\cal D}}

\newcommand {\cL}{{\cal L}}
\newcommand {\cM}{{\cal M}}

\def\a{\alpha}
\def\b{\beta}
\def\c{\chi}
\def\d{\delta}

\def\f{\phi}

\def\G{\Gamma}

\def\k{\kappa}
\def\l{\lambda}
\def\m{\mu}
\def\n{\nu}
\def\o{\omega}

\def\q{\theta}
\def\r{\rho}
\def\s{\sigma}
\def\t{\tau}

\def\x{\xi}
\def\z{\zeta}

\def\F{\Phi}
\def\J{\Psi}
\def\L{\Lambda}
\def\O{\Omega}

\newcommand{\ad}{{\dot{\alpha}}}                           
\newcommand{\ve}{\varepsilon}                            

\newcommand{\hf}{\frac12}

\newcommand{\be}{\begin{equation}}
\newcommand{\ee}{\end{equation}}
\newcommand{\bea}{\begin{eqnarray}}
\newcommand{\eea}{\end{eqnarray}}
\newcommand{\non}{\nonumber}
\begin{document}
\immediate\write16{<WARNING: FEYNMAN macros work only with emTeX-dvivers
                    (dviscr.exe, dvihplj.exe, dvidot.exe, etc.) >}
\newdimen\Lengthunit
\newcount\Nhalfperiods
\Lengthunit = 1.5cm
\Nhalfperiods = 9
\catcode`\*=11
\newdimen\L*   \newdimen\d*   \newdimen\d**
\newdimen\dm*  \newdimen\dd*  \newdimen\dt*
\newdimen\a*   \newdimen\b*   \newdimen\c*
\newdimen\a**  \newdimen\b**
\newdimen\xL*  \newdimen\yL*
\newcount\k*   \newcount\l*   \newcount\m*
\newcount\n*   \newcount\dn*  \newcount\r*
\newcount\N*   \newcount\*one \newcount\*two  \*one=1 \*two=2
\newcount\*ths \*ths=1000
\def\GRAPH(hsize=#1)#2{\hbox to #1\Lengthunit{#2\hss}}
\def\Linewidth#1{\special{em:linewidth #1}}
\Linewidth{.4pt}
\def\sm*{\special{em:moveto}}
\def\sl*{\special{em:lineto}}
\newbox\spm*   \newbox\spl*
\setbox\spm*\hbox{\sm*}
\setbox\spl*\hbox{\sl*}
\def\mov#1(#2,#3)#4{\rlap{\L*=#1\Lengthunit\kern#2\L*\raise#3\L*\hbox{#4}}}
\def\smov#1(#2,#3)#4{\rlap{\L*=#1\Lengthunit
\xL*=\xscale\L*\yL*=\yscale\L*\kern#2\xL*\raise#3\yL*\hbox{#4}}}
\def\mov*(#1,#2)#3{\rlap{\kern#1\raise#2\hbox{#3}}}
\def\lin#1(#2,#3){\rlap{\sm*\mov#1(#2,#3){\sl*}}}
\def\arr*(#1,#2,#3){\mov*(#1\dd*,#1\dt*){%
\sm*\mov*(#2\dd*,#2\dt*){\mov*(#3\dt*,-#3\dd*){\sl*}}%
\sm*\mov*(#2\dd*,#2\dt*){\mov*(-#3\dt*,#3\dd*){\sl*}}}}
\def\arrow#1(#2,#3){\rlap{\lin#1(#2,#3)\mov#1(#2,#3){%
\d**=-.012\Lengthunit\dd*=#2\d**\dt*=#3\d**%
\arr*(1,10,4)\arr*(3,8,4)\arr*(4.8,4.2,3)}}}
\def\arrlin#1(#2,#3){\rlap{\L*=#1\Lengthunit\L*=.5\L*%
\lin#1(#2,#3)\mov*(#2\L*,#3\L*){\arrow.1(#2,#3)}}}
\def\dasharrow#1(#2,#3){\rlap{%
{\Lengthunit=0.9\Lengthunit\dashlin#1(#2,#3)\mov#1(#2,#3){\sm*}}%
\mov#1(#2,#3){\sl*\d**=-.012\Lengthunit\dd*=#2\d**\dt*=#3\d**%
\arr*(1,10,4)\arr*(3,8,4)\arr*(4.8,4.2,3)}}}
\def\clap#1{\hbox to 0pt{\hss #1\hss}}
\def\ind(#1,#2)#3{\rlap{%
\d*=.1\Lengthunit\kern#1\d*\raise#2\d*\hbox{\lower2pt\clap{$#3$}}}}
\def\sh*(#1,#2)#3{\rlap{%
\dm*=\the\n*\d**\xL*=\xscale\dm*\yL*=\yscale\dm*
\kern#1\xL*\raise#2\yL*\hbox{#3}}}
\def\calcnum*#1(#2,#3){\a*=1000sp\b*=1000sp\a*=#2\a*\b*=#3\b*%
\ifdim\a*<0pt\a*-\a*\fi\ifdim\b*<0pt\b*-\b*\fi%
\ifdim\a*>\b*\c*=.96\a*\advance\c*.4\b*%
\else\c*=.96\b*\advance\c*.4\a*\fi%
\k*\a*\multiply\k*\k*\l*\b*\multiply\l*\l*%
\m*\k*\advance\m*\l*\n*\c*\r*\n*\multiply\n*\n*%
\dn*\m*\advance\dn*-\n*\divide\dn*2\divide\dn*\r*%
\advance\r*\dn*%
\c*=\the\Nhalfperiods5sp\c*=#1\c*\ifdim\c*<0pt\c*-\c*\fi%
\multiply\c*\r*\N*\c*\divide\N*10000}
\def\dashlin#1(#2,#3){\rlap{\calcnum*#1(#2,#3)%
\d**=#1\Lengthunit\ifdim\d**<0pt\d**-\d**\fi%
\divide\N*2\multiply\N*2\advance\N*1%
\divide\d**\N*\sm*\n*\*one\sh*(#2,#3){\sl*}%
\loop\advance\n*\*one\sh*(#2,#3){\sm*}\advance\n*\*one\sh*(#2,#3){\sl*}%
\ifnum\n*<\N*\repeat}}
\def\dashdotlin#1(#2,#3){\rlap{\calcnum*#1(#2,#3)%
\d**=#1\Lengthunit\ifdim\d**<0pt\d**-\d**\fi%
\divide\N*2\multiply\N*2\advance\N*1\multiply\N*2%
\divide\d**\N*\sm*\n*\*two\sh*(#2,#3){\sl*}\loop%
\advance\n*\*one\sh*(#2,#3){\kern-1.48pt\lower.5pt\hbox{\rm.}}%
\advance\n*\*one\sh*(#2,#3){\sm*}%
\advance\n*\*two\sh*(#2,#3){\sl*}\ifnum\n*<\N*\repeat}}
\def\shl*(#1,#2)#3{\kern#1#3\lower#2#3\hbox{\unhcopy\spl*}}
\def\trianglin#1(#2,#3){\rlap{\toks0={#2}\toks1={#3}\calcnum*#1(#2,#3)%
\dd*=.57\Lengthunit\dd*=#1\dd*\divide\dd*\N*%
\d**=#1\Lengthunit\ifdim\d**<0pt\d**-\d**\fi%
\multiply\N*2\divide\d**\N*\advance\N*-1\sm*\n*\*one\loop%
\shl**{\dd*}\dd*-\dd*\advance\n*2%
\ifnum\n*<\N*\repeat\n*\N*\advance\n*1\shl**{0pt}}}
\def\wavelin#1(#2,#3){\rlap{\toks0={#2}\toks1={#3}\calcnum*#1(#2,#3)%
\dd*=.23\Lengthunit\dd*=#1\dd*\divide\dd*\N*%
\d**=#1\Lengthunit\ifdim\d**<0pt\d**-\d**\fi%
\multiply\N*4\divide\d**\N*\sm*\n*\*one\loop%
\shl**{\dd*}\dt*=1.3\dd*\advance\n*1%
\shl**{\dt*}\advance\n*\*one%
\shl**{\dd*}\advance\n*\*two%
\dd*-\dd*\ifnum\n*<\N*\repeat\n*\N*\shl**{0pt}}}
\def\w*lin(#1,#2){\rlap{\toks0={#1}\toks1={#2}\d**=\Lengthunit\dd*=-.12\d**%
\N*8\divide\d**\N*\sm*\n*\*one\loop%
\shl**{\dd*}\dt*=1.3\dd*\advance\n*\*one%
\shl**{\dt*}\advance\n*\*one%
\shl**{\dd*}\advance\n*\*one%
\shl**{0pt}\dd*-\dd*\advance\n*1\ifnum\n*<\N*\repeat}}
\def\l*arc(#1,#2)[#3][#4]{\rlap{\toks0={#1}\toks1={#2}\d**=\Lengthunit%
\dd*=#3.037\d**\dd*=#4\dd*\dt*=#3.049\d**\dt*=#4\dt*\ifdim\d**>16mm%
\d**=.25\d**\n*\*one\shl**{-\dd*}\n*\*two\shl**{-\dt*}\n*3\relax%
\shl**{-\dd*}\n*4\relax\shl**{0pt}\else\ifdim\d**>5mm%
\d**=.5\d**\n*\*one\shl**{-\dt*}\n*\*two\shl**{0pt}%
\else\n*\*one\shl**{0pt}\fi\fi}}
\def\d*arc(#1,#2)[#3][#4]{\rlap{\toks0={#1}\toks1={#2}\d**=\Lengthunit%
\dd*=#3.037\d**\dd*=#4\dd*\d**=.25\d**\sm*\n*\*one\shl**{-\dd*}%
\n*3\relax\sh*(#1,#2){\xL*=\xscale\dd*\yL*=\yscale\dd*
\kern#2\xL*\lower#1\yL*\hbox{\sm*}}%
\n*4\relax\shl**{0pt}}}
\def\arc#1[#2][#3]{\rlap{\Lengthunit=#1\Lengthunit%
\sm*\l*arc(#2.1914,#3.0381)[#2][#3]%
\smov(#2.1914,#3.0381){\l*arc(#2.1622,#3.1084)[#2][#3]}%
\smov(#2.3536,#3.1465){\l*arc(#2.1084,#3.1622)[#2][#3]}%
\smov(#2.4619,#3.3086){\l*arc(#2.0381,#3.1914)[#2][#3]}}}
\def\dasharc#1[#2][#3]{\rlap{\Lengthunit=#1\Lengthunit%
\d*arc(#2.1914,#3.0381)[#2][#3]%
\smov(#2.1914,#3.0381){\d*arc(#2.1622,#3.1084)[#2][#3]}%
\smov(#2.3536,#3.1465){\d*arc(#2.1084,#3.1622)[#2][#3]}%
\smov(#2.4619,#3.3086){\d*arc(#2.0381,#3.1914)[#2][#3]}}}
\def\wavearc#1[#2][#3]{\rlap{\Lengthunit=#1\Lengthunit%
\w*lin(#2.1914,#3.0381)%
\smov(#2.1914,#3.0381){\w*lin(#2.1622,#3.1084)}%
\smov(#2.3536,#3.1465){\w*lin(#2.1084,#3.1622)}%
\smov(#2.4619,#3.3086){\w*lin(#2.0381,#3.1914)}}}
\def\shl**#1{\c*=\the\n*\d**\d*=#1%
\a*=\the\toks0\c*\b*=\the\toks1\d*\advance\a*-\b*%
\b*=\the\toks1\c*\d*=\the\toks0\d*\advance\b*\d*%
\a*=\xscale\a*\b*=\yscale\b*%
\raise\b*\rlap{\kern\a*\unhcopy\spl*}}
\def\wlin*#1(#2,#3)[#4]{\rlap{\toks0={#2}\toks1={#3}%
\c*=#1\l*\c*\c*=.01\Lengthunit\m*\c*\divide\l*\m*%
\c*=\the\Nhalfperiods5sp\multiply\c*\l*\N*\c*\divide\N*\*ths%
\divide\N*2\multiply\N*2\advance\N*1%
\dd*=.002\Lengthunit\dd*=#4\dd*\multiply\dd*\l*\divide\dd*\N*%
\d**=#1\multiply\N*4\divide\d**\N*\sm*\n*\*one\loop%
\shl**{\dd*}\dt*=1.3\dd*\advance\n*\*one%
\shl**{\dt*}\advance\n*\*one%
\shl**{\dd*}\advance\n*\*two%
\dd*-\dd*\ifnum\n*<\N*\repeat\n*\N*\shl**{0pt}}}
\def\wavebox#1{\setbox0\hbox{#1}%
\a*=\wd0\advance\a*14pt\b*=\ht0\advance\b*\dp0\advance\b*14pt%
\hbox{\kern9pt%
\mov*(0pt,\ht0){\mov*(-7pt,7pt){\wlin*\a*(1,0)[+]\wlin*\b*(0,-1)[-]}}%
\mov*(\wd0,-\dp0){\mov*(7pt,-7pt){\wlin*\a*(-1,0)[+]\wlin*\b*(0,1)[-]}}%
\box0\kern9pt}}
\def\rectangle#1(#2,#3){%
\lin#1(#2,0)\lin#1(0,#3)\mov#1(0,#3){\lin#1(#2,0)}\mov#1(#2,0){\lin#1(0,#3)}}
\def\dashrectangle#1(#2,#3){\dashlin#1(#2,0)\dashlin#1(0,#3)%
\mov#1(0,#3){\dashlin#1(#2,0)}\mov#1(#2,0){\dashlin#1(0,#3)}}
\def\waverectangle#1(#2,#3){\L*=#1\Lengthunit\a*=#2\L*\b*=#3\L*%
\ifdim\a*<0pt\a*-\a*\def\x*{-1}\else\def\x*{1}\fi%
\ifdim\b*<0pt\b*-\b*\def\y*{-1}\else\def\y*{1}\fi%
\wlin*\a*(\x*,0)[-]\wlin*\b*(0,\y*)[+]%
\mov#1(0,#3){\wlin*\a*(\x*,0)[+]}\mov#1(#2,0){\wlin*\b*(0,\y*)[-]}}
\def\calcparab*{%
\ifnum\n*>\m*\k*\N*\advance\k*-\n*\else\k*\n*\fi%
\a*=\the\k* sp\a*=10\a*\b*\dm*\advance\b*-\a*\k*\b*%
\a*=\the\*ths\b*\divide\a*\l*\multiply\a*\k*%
\divide\a*\l*\k*\*ths\r*\a*\advance\k*-\r*%
\dt*=\the\k*\L*}
\def\arcto#1(#2,#3)[#4]{\rlap{\toks0={#2}\toks1={#3}\calcnum*#1(#2,#3)%
\dm*=135sp\dm*=#1\dm*\d**=#1\Lengthunit\ifdim\dm*<0pt\dm*-\dm*\fi%
\multiply\dm*\r*\a*=.3\dm*\a*=#4\a*\ifdim\a*<0pt\a*-\a*\fi%
\advance\dm*\a*\N*\dm*\divide\N*10000%
\divide\N*2\multiply\N*2\advance\N*1%
\L*=-.25\d**\L*=#4\L*\divide\d**\N*\divide\L*\*ths%
\m*\N*\divide\m*2\dm*=\the\m*5sp\l*\dm*%
\sm*\n*\*one\loop\calcparab*\shl**{-\dt*}%
\advance\n*1\ifnum\n*<\N*\repeat}}
\def\arrarcto#1(#2,#3)[#4]{\L*=#1\Lengthunit\L*=.54\L*%
\arcto#1(#2,#3)[#4]\mov*(#2\L*,#3\L*){\d*=.457\L*\d*=#4\d*\d**-\d*%
\mov*(#3\d**,#2\d*){\arrow.02(#2,#3)}}}
\def\dasharcto#1(#2,#3)[#4]{\rlap{\toks0={#2}\toks1={#3}\calcnum*#1(#2,#3)%
\dm*=\the\N*5sp\a*=.3\dm*\a*=#4\a*\ifdim\a*<0pt\a*-\a*\fi%
\advance\dm*\a*\N*\dm*%
\divide\N*20\multiply\N*2\advance\N*1\d**=#1\Lengthunit%
\L*=-.25\d**\L*=#4\L*\divide\d**\N*\divide\L*\*ths%
\m*\N*\divide\m*2\dm*=\the\m*5sp\l*\dm*%
\sm*\n*\*one\loop%
\calcparab*\shl**{-\dt*}\advance\n*1%
\ifnum\n*>\N*\else\calcparab*%
\sh*(#2,#3){\kern#3\dt*\lower#2\dt*\hbox{\sm*}}\fi%
\advance\n*1\ifnum\n*<\N*\repeat}}
\def\*shl*#1{%
\c*=\the\n*\d**\advance\c*#1\a**\d*\dt*\advance\d*#1\b**%
\a*=\the\toks0\c*\b*=\the\toks1\d*\advance\a*-\b*%
\b*=\the\toks1\c*\d*=\the\toks0\d*\advance\b*\d*%
\raise\b*\rlap{\kern\a*\unhcopy\spl*}}
\def\calcnormal*#1{%
\b**=10000sp\a**\b**\k*\n*\advance\k*-\m*%
\multiply\a**\k*\divide\a**\m*\a**=#1\a**\ifdim\a**<0pt\a**-\a**\fi%
\ifdim\a**>\b**\d*=.96\a**\advance\d*.4\b**%
\else\d*=.96\b**\advance\d*.4\a**\fi%
\d*=.01\d*\r*\d*\divide\a**\r*\divide\b**\r*%
\ifnum\k*<0\a**-\a**\fi\d*=#1\d*\ifdim\d*<0pt\b**-\b**\fi%
\k*\a**\a**=\the\k*\dd*\k*\b**\b**=\the\k*\dd*}
\def\wavearcto#1(#2,#3)[#4]{\rlap{\toks0={#2}\toks1={#3}\calcnum*#1(#2,#3)%
\c*=\the\N*5sp\a*=.4\c*\a*=#4\a*\ifdim\a*<0pt\a*-\a*\fi%
\advance\c*\a*\N*\c*\divide\N*20\multiply\N*2\advance\N*-1\multiply\N*4%
\d**=#1\Lengthunit\dd*=.012\d**\ifdim\d**<0pt\d**-\d**\fi\L*=.25\d**%
\divide\d**\N*\divide\dd*\N*\L*=#4\L*\divide\L*\*ths%
\m*\N*\divide\m*2\dm*=\the\m*0sp\l*\dm*%
\sm*\n*\*one\loop\calcnormal*{#4}\calcparab*%
\*shl*{1}\advance\n*\*one\calcparab*%
\*shl*{1.3}\advance\n*\*one\calcparab*%
\*shl*{1}\advance\n*2%
\dd*-\dd*\ifnum\n*<\N*\repeat\n*\N*\shl**{0pt}}}
\def\triangarcto#1(#2,#3)[#4]{\rlap{\toks0={#2}\toks1={#3}\calcnum*#1(#2,#3)%
\c*=\the\N*5sp\a*=.4\c*\a*=#4\a*\ifdim\a*<0pt\a*-\a*\fi%
\advance\c*\a*\N*\c*\divide\N*20\multiply\N*2\advance\N*-1\multiply\N*2%
\d**=#1\Lengthunit\dd*=.012\d**\ifdim\d**<0pt\d**-\d**\fi\L*=.25\d**%
\divide\d**\N*\divide\dd*\N*\L*=#4\L*\divide\L*\*ths%
\m*\N*\divide\m*2\dm*=\the\m*0sp\l*\dm*%
\sm*\n*\*one\loop\calcnormal*{#4}\calcparab*%
\*shl*{1}\advance\n*2%
\dd*-\dd*\ifnum\n*<\N*\repeat\n*\N*\shl**{0pt}}}
\def\hr*#1{\clap{\xL*=\xscale\Lengthunit\vrule width#1\xL* height.1pt}}
\def\shade#1[#2]{\rlap{\Lengthunit=#1\Lengthunit%
\smov(0,#2.05){\hr*{.994}}\smov(0,#2.1){\hr*{.980}}%
\smov(0,#2.15){\hr*{.953}}\smov(0,#2.2){\hr*{.916}}%
\smov(0,#2.25){\hr*{.867}}\smov(0,#2.3){\hr*{.798}}%
\smov(0,#2.35){\hr*{.715}}\smov(0,#2.4){\hr*{.603}}%
\smov(0,#2.45){\hr*{.435}}}}
\def\dshade#1[#2]{\rlap{%
\Lengthunit=#1\Lengthunit\if#2-\def\t*{+}\else\def\t*{-}\fi%
\smov(0,\t*.025){%
\smov(0,#2.05){\hr*{.995}}\smov(0,#2.1){\hr*{.988}}%
\smov(0,#2.15){\hr*{.969}}\smov(0,#2.2){\hr*{.937}}%
\smov(0,#2.25){\hr*{.893}}\smov(0,#2.3){\hr*{.836}}%
\smov(0,#2.35){\hr*{.760}}\smov(0,#2.4){\hr*{.662}}%
\smov(0,#2.45){\hr*{.531}}\smov(0,#2.5){\hr*{.320}}}}}
\def\vdot{\rlap{\kern-1.9pt\lower1.8pt\hbox{$\scriptstyle\bullet$}}}
\def\vtimes{\rlap{\kern-3pt\lower1.8pt\hbox{$\scriptstyle\times$}}}
\def\vDot{\rlap{\kern-2.3pt\lower2.7pt\hbox{$\bullet$}}}
\def\vTimes{\rlap{\kern-3.6pt\lower2.4pt\hbox{$\times$}}}
\catcode`\*=12
\newcount\CatcodeOfAtSign
\CatcodeOfAtSign=\the\catcode`\@
\catcode`\@=11
\newcount\n@ast
\def\n@ast@#1{\n@ast0\relax\get@ast@#1\end}
\def\get@ast@#1{\ifx#1\end\let\next\relax\else%
\ifx#1*\advance\n@ast1\fi\let\next\get@ast@\fi\next}
\newif\if@up \newif\if@dwn
\def\up@down@#1{\@upfalse\@dwnfalse%
\if#1u\@uptrue\fi\if#1U\@uptrue\fi\if#1+\@uptrue\fi%
\if#1d\@dwntrue\fi\if#1D\@dwntrue\fi\if#1-\@dwntrue\fi}
\def\halfcirc#1(#2)[#3]{{\Lengthunit=#2\Lengthunit\up@down@{#3}%
\if@up\smov(0,.5){\arc[-][-]\arc[+][-]}\fi%
\if@dwn\smov(0,-.5){\arc[-][+]\arc[+][+]}\fi%
\def\lft{\smov(0,.5){\arc[-][-]}\smov(0,-.5){\arc[-][+]}}%
\def\rght{\smov(0,.5){\arc[+][-]}\smov(0,-.5){\arc[+][+]}}%
\if#3l\lft\fi\if#3L\lft\fi\if#3r\rght\fi\if#3R\rght\fi%
\n@ast@{#1}%
\ifnum\n@ast>0\if@up\shade[+]\fi\if@dwn\shade[-]\fi\fi%
\ifnum\n@ast>1\if@up\dshade[+]\fi\if@dwn\dshade[-]\fi\fi}}
\def\halfdashcirc(#1)[#2]{{\Lengthunit=#1\Lengthunit\up@down@{#2}%
\if@up\smov(0,.5){\dasharc[-][-]\dasharc[+][-]}\fi%
\if@dwn\smov(0,-.5){\dasharc[-][+]\dasharc[+][+]}\fi%
\def\lft{\smov(0,.5){\dasharc[-][-]}\smov(0,-.5){\dasharc[-][+]}}%
\def\rght{\smov(0,.5){\dasharc[+][-]}\smov(0,-.5){\dasharc[+][+]}}%
\if#2l\lft\fi\if#2L\lft\fi\if#2r\rght\fi\if#2R\rght\fi}}
\def\halfwavecirc(#1)[#2]{{\Lengthunit=#1\Lengthunit\up@down@{#2}%
\if@up\smov(0,.5){\wavearc[-][-]\wavearc[+][-]}\fi%
\if@dwn\smov(0,-.5){\wavearc[-][+]\wavearc[+][+]}\fi%
\def\lft{\smov(0,.5){\wavearc[-][-]}\smov(0,-.5){\wavearc[-][+]}}%
\def\rght{\smov(0,.5){\wavearc[+][-]}\smov(0,-.5){\wavearc[+][+]}}%
\if#2l\lft\fi\if#2L\lft\fi\if#2r\rght\fi\if#2R\rght\fi}}
\def\Circle#1(#2){\halfcirc#1(#2)[u]\halfcirc#1(#2)[d]\n@ast@{#1}%
\ifnum\n@ast>0\clap{%
\dimen0=\xscale\Lengthunit\vrule width#2\dimen0 height.1pt}\fi}
\def\wavecirc(#1){\halfwavecirc(#1)[u]\halfwavecirc(#1)[d]}
\def\dashcirc(#1){\halfdashcirc(#1)[u]\halfdashcirc(#1)[d]}
%
\def\xscale{1}
\def\yscale{1}
\def\Ellipse#1(#2)[#3,#4]{\def\xscale{#3}\def\yscale{#4}%
\Circle#1(#2)\def\xscale{1}\def\yscale{1}}
\def\dashEllipse(#1)[#2,#3]{\def\xscale{#2}\def\yscale{#3}%
\dashcirc(#1)\def\xscale{1}\def\yscale{1}}
\def\waveEllipse(#1)[#2,#3]{\def\xscale{#2}\def\yscale{#3}%
\wavecirc(#1)\def\xscale{1}\def\yscale{1}}
\def\halfEllipse#1(#2)[#3][#4,#5]{\def\xscale{#4}\def\yscale{#5}%
\halfcirc#1(#2)[#3]\def\xscale{1}\def\yscale{1}}
\def\halfdashEllipse(#1)[#2][#3,#4]{\def\xscale{#3}\def\yscale{#4}%
\halfdashcirc(#1)[#2]\def\xscale{1}\def\yscale{1}}
\def\halfwaveEllipse(#1)[#2][#3,#4]{\def\xscale{#3}\def\yscale{#4}%
\halfwavecirc(#1)[#2]\def\xscale{1}\def\yscale{1}}
\catcode`\@=\the\CatcodeOfAtSign

\begin{titlepage}
\thispagestyle{empty}

\begin{flushright}
UPR-775T\\
IASSNS-97/109\\
ITP-UH-25/97 \\
hep-th/9710142 \\
\end{flushright}

\begin{center}
{\large\bf On the D = 4, N = 2  Non-Renormalization Theorem
}
\end{center}

\begin{center}
{\bf Ioseph L. Buchbinder}\\
\footnotesize{
{\it
Department of Theoretical Physics,
Tomsk State Pedagogical University\\
Tomsk 634041, Russia}
}
\end{center}

\begin{center}
{\bf Sergei M. Kuzenko}\footnote{Alexander von Humboldt Research
Fellow. On leave from Department of Quantum Field Theory,
Tomsk State University, Tomsk 634050, Russia.}\\
\footnotesize{
{\it Institut f\"ur Theoretische Physik,
Universit\"{a}t Hannover\\
Appelstra{\ss}e 2, D-30167 Hannover, Germany}
}
\end{center}

\begin{center}
{\bf Burt A. Ovrut} \\
\footnotesize{
{\it
Department of Physics, University of Pennsylvania\\
Philadelphia, PA 19104-6396, USA}}\\
and\\
\footnotesize{
{\it Institut f\"ur Physik, Humboldt-Universit\"at zu Berlin\\
Invalidenstra{\ss}e 110, D-10115 Berlin, Germany}}\\
and\\
\footnotesize{
{\it School of Natural Sciences, Institute for Advanced Study\\
Olden Lane, Princeton, NJ 08540, USA}}
\end{center}

\begin{abstract}
Using the harmonic superspace background field formulation for
general $D=4$, $N=2$ super Yang-Mills theories, with matter
hypermultiplets in arbitrary representations of the gauge group,
we present the first rigorous proof of the $N=2$ non-renormalization
theorem; specifically, the
absence of ultraviolet divergences beyond the one-loop level.
Another simple consequence of the background field formulation is
the absence of the leading non-holomorphic correction
to the low-energy effective action at two loops.
\vfill
\end{abstract}
\end{titlepage}

\newpage
\setcounter{page}{1}
\noindent
There are two basic formulations of the $N=2$, $D=4$ pure super
Yang-Mills theory in terms of unconstrained superfields. The first
(conventional) formulation,
which was developed at the linearized level by Mezincescu \cite{m}
and then extended to the full nonlinear theory by Koller and Howe,
Stelle and Townsend
\cite{k,hst}, makes use of the conventional $N=2$ superspace
${\bf R}^{4|8}$ parametrized by
$z^M\equiv(x^m,\theta_i^\alpha, {\bar\theta}^i_{\dot\alpha})$ where
${\bar \q}^{i \ad} = \overline{\q^\a_i}$.
The unconstrained prepotential of this theory, $U^{ij}(z)$, is an
isovector real superfield,
$U^{ij}= U^{(ij)} = \overline{U_{ij}}$,
taking its values in the Lie algebra of the gauge group. In this approach,
the $N=2$ super Yang-Mills theory possesses a non-trivial gauge invariance
with an infinite degree of reducibility \cite{sg}.
The second (harmonic) formulation,
developed by GIKOS \cite{gikos}, makes use of the
$N=2$ harmonic superspace ${\bf R}^{4|8}\times S^2$. This approach
extends the conventional superspace by the
two-sphere $S^2 =SU(2)/U(1)$
parametrized by harmonics; that is, group
elements
\bea
&({u_i}^-\,,\,{u_i}^+) \in SU(2)\non\\
&u^+_i = \ve_{ij}u^{+j} \qquad \overline{u^{+i}} = u^-_i
\qquad u^{+i}u_i^- = 1 \;.
\label{2}
\eea
The unconstrained prepotential of this theory is an analytic real
Lie-algebra valued superfield $V^{++}(\z, u)$. This superfield
is defined over the analytic
subspace of the harmonic superspace parametrized by the variables
\be
\z^\cM \equiv (x^m_A,\q^{+\a},{\bar\q}^+_{\dot\a}, \,
u^+_i,u^-_j)
\label{3}
\ee
where the analytic basis in the subspace is defined by
\be
x^m_A = x^m - 2{\rm i} \q^{(i}\s^m {\bar \q}^{j)}u^+_i u^-_j \qquad
 \q^\pm_\a=u^\pm_i \q^i_\a \qquad {\bar \q}^\pm_{\dot\a}=u^\pm_i{\bar
\q}^i_{\dot\a}\;.
\label{4}
\ee
In this approach the $N=2$ super Yang-Mills theory is an irreducible
gauge theory.

Because of the infinitely reducible gauge structure of $N=2$
super Yang-Mills theory formulated in conventional superspace,
its quantization cannot be carried out using the Faddeev-Popov prescription
and should be based on more powerful quantization techniques,
such as the Batalin-Vilkovisky method \cite{bv} (even this latter scheme is
literally applicable to finitely reducible gauge theories only).
To the best of our knowledge, the Batalin-Vilkovisky quantization
of the theory has never been utilized, in this context, to derive a consistent
superfield effective action. Instead, the two attempts
to quantize this theory, undertaken in \cite{hst,mar}, were based on a
modified Faddeev-Popov prescription, which has not been shown to be a
consistent quantization scheme.
Furthermore, although the $N=2$ background field method
presented in \cite{hst}
has played a significant role in understanding the general structure
of extended supersymmetric theories, this approach is very complicated
from the technical point of view and is not suitable for carrying out
actual quantum computations.
It is disturbing, therefore, that the original proof of
the famous $N=2$ non-renormalization theorem
(see, for example, \cite{ggrs,west} and references therein) assumes not only
the existence of an unconstrained classical formulation in conventional
superspace, but also a consistent formulation of the superfield Feynman
rules in this superspace which, as we have seen, has yet to be developed.
An indirect proof of the $N=2$ non-renormalization theorem,
based on an explicit calculation of the one-loop $N=2$ beta function 
and the application of anomalies considerations, was presented in \cite{beta}.
A different approach to quantum calculations in $N=2$ supersymmetric theories
is to reformulate them in terms of  $N=1$ superspace \cite{n1cal}, and then
to use the usual $N=1$ supergraph techniques
or instanton methods. Here, too, there are
fundamental problems. To begin with, in this approach, the second
supersymmetry is hidden. More importantly, it is far from clear that
the regulators used
in this approach respect the $N=2$ supersymmetry. Hence, quantum
corrected Greens functions may not necessarily be $N=2$ supersymmetric.
It follows that, at the very least, the inherent mechanisms of the miraculous
cancellations of ultraviolet divergences are not manifest. It has also yet to
be proven that this technique preserves $N=2$ supersymmetry to to all loop levels
in quantum corrections.
We conclude that
the $N=2$ non-renormalization theorem requires more careful
justification than has previously appeared in the literature.
Recently, the first examples of quantum calculations with manifest
$N=2$ supersymmetry
have been given within the context of harmonic superspace
\cite{bbiko,ikz}. In this paper, we will use these new techniques to give a
rigorous proof of the $N=2$ non-renormalization theorem, as well as to
establish the absence of the leading finite non-holomorphic correction at the
two-loop level.

It has been known for a long time \cite{gikos,giosreview}
that the conventional superfield
formulation of the $N=2$ super Yang-Mills theory
is simply a gauge fixed version of that theory in the
harmonic superspace. More precisely, if one expresses the analytic
prepotential $V^{++}(\z, u)$ in terms of an unconstrained superfield
$U^{--}(z, u)$ over ${\bf R}^{4|8} \times S^2$
(and similarly for the analytic gauge parameter)
\bea
V^{++} (\z, u) &=& (D^+)^4 \,U^{--} (z, u) \non \\
U^{--} (z,u) &=& U^{(ij)} (z)\, u^-_i u^-_j +
U^{(ijkl)} (z) \,u^+_i u^+_j u^-_k u^-_l + \cdots
\label{repar}
\eea
then the original gauge freedom can be used to gauge away all but the
$U^{ij}(z)$ components of $U^{--}(z,u)$; the remaining superfield
$U^{ij} (z)$ being exactly Mezincescu's prepotenial.
Since the harmonic formulation
of the $N=2$ super Yang-Mills theory is an irreducible gauge theory,
it can, unlike the conventional formulation, be properly
quantized using the standard
Faddeev-Popov prescription \cite{gios}. In the harmonic formulation,
we simply have none of the quantization problems that are
inevitable in the conventional
superspace approach. Moreover, harmonic superspace
allows us to describe matter hypermultiplets in arbitrary representations
of the gauge group in terms of unconstrained analytic superfields
\cite{gikos,hsw}. The above remarkable features make the harmonic formulation
unique and, in principle, indispensable for the study of the quantum aspects
of $N=2$ super Yang-Mills theories.

In a recent paper \cite{back}, we have presented the background field
method for general $N=2$ super Yang-Mills theories in harmonic superspace.
The purpose of this paper is to show that
this method makes it possible to develop a covariant $N=2$
diagram technique, very much like the well known $N=1$ supergraph techniques
(see \cite{ggrs,west,bk} for a review), and, for the first time,
to rigorously prove the $N=2$
non-renormalization theorem. In addition, the harmonic superspace
background field method allows us to obtain some important results
concerning the finite structure of the low-energy effective action at higher
loops.

The harmonic formulation is naturally compatible with two pictures used
to describe the $N=2$ gauge supermultiplet \cite{gikos}, and they prove
to be very useful both at the classical and quantum levels.
In the first picture, called the $\t$-frame, the connection is
$u$-independent. The gauge covariant derivatives read
\bea
& {\cal D}_{\underline{M}}\equiv({\cal D}_M,D^{++},D^{--},
D^0) \non \\
& {\cal D}_M\equiv ({\cal D}_m,{\cal
D}^i_\alpha,{\bar{\cal D}}_i^{\dot\alpha})= D_M+{\rm i}A_M \qquad
A_M=A_M^a(z)T^a
\label{5}
\eea
and satisfy the algebra
\bea
& \{{\cal D}^i_\alpha,{\bar{\cal D}}_{{\dot\alpha j}}\}= -2{\rm i}\delta^i_j
{\cal D}_{\alpha{\dot\alpha}} \non \\
&\{{\cal D}^i_\alpha,{\cal D}^j_\beta\}=2{\rm i}\varepsilon_{\alpha\beta}
\varepsilon^{ij}{\bar W}\qquad \{{\bar{\cal D}}_{{\dot\alpha}i},
{\bar{\cal D}}_{{\dot\beta}j}\}=2{\rm
i}\varepsilon_{{\dot\alpha}{\dot\beta}}
\varepsilon_{ij}W
\label{6}\\
&{[}D^{\pm \pm},{\cal D}_M{]} = {[}D^0, {\cal D}_M] = 0\;. \non
\eea
Here $D_M\equiv(\partial_m,D^i_\alpha,{\bar D}^{\dot\alpha}_i)$ are the
flat covariant derivatives, $T^a$ the generators of the gauge group
and the harmonic derivatives look like \cite{gios}
\be
D^{\pm\pm}=u^{\pm i}\frac{\partial}{\partial u^{\mp i}} \qquad
D^0=u^{+i}\frac{\partial}{\partial u^{+i}}-u^{-i}
\frac{\partial}{\partial u^{-i}} \;.
\label{7}
\ee
The covariant derivatives and a matter superfield multiplet $\J(z,u)$
transform under the gauge group as follows
\be
{\cal D}'_{\underline{M}}=e^{{\rm i}\tau}
{\cal D}_{\underline{M}} e^{-{\rm i}\tau}
\qquad \J'=e^{{\rm i}\tau}\J \qquad \tau=\tau^a(z)T^a
\label{8}
\ee
with $\tau^a$ being real $u$-independent unconstrained parameters.
The existence of the second picture, called the $\l$-frame,
follows from the algebra (\ref{6}). Introducing
${\cal D}^\pm_\alpha=u^\pm_i{\cal
D}^i_\alpha$ and ${\bar{\cal D}}^\pm_{\dot\alpha}=u^\pm_i{\bar{\cal
D}}^i_{\dot\alpha}$, one observes that the operators
$\cD^+_\a$ and ${\bar \cD}^+_\ad$ anticommute. Hence
\be
{\cal D}^+_\alpha=e^{-{\rm i}\Omega}D^+_\alpha e^{{\rm i}\Omega} \qquad
{\bar{\cal D}}^+_{\dot\alpha}=e^{-{\rm i}\Omega}{\bar D}^+_{\dot\alpha}
e^{{\rm i}\Omega} \qquad \Omega=\Omega^a(z,u)T^a
\label{9}
\ee
for some Lie-algebra valued superfield $\Omega=\Omega^a(z,u)T^a$,
called the bridge. Superfield $\Omega$ has
zero $U(1)$-charge, $D^0\Omega^a=0$, and is real,
${\stackrel{\smile}{\Omega}}{}^a = \Omega^a$,
with respect to the analyticity-preserving conjugation \cite{gikos},
which we denote here by $\smile$. As a consequnce, one can define new
superfield types; that is,
covariantly analytic superfields constrained by
\be
{\cal D}^+_\alpha\Phi^{(q)}={\bar{\cal D}}^+_{\dot\alpha} \Phi^{(q)}=0\;.
\label{10}
\ee
Here $\Phi^{(q)}(z,u)$ carries $U(1)$-charge $q$, $D^0\Phi^{(q)}=q\Phi^
{(q)}$, and can be represented as follows
\be
\Phi^{(q)}=e^{-{\rm i}\Omega}\phi^{(q)} \qquad
D^+_\alpha\phi^{(q)}={\bar D}^+_{\dot\alpha}\phi^{(q)}=0
\label{11}
\ee
with $\phi^{(q)}(\zeta, u)$ being
an unconstrained superfield over the analytic subspace (\ref{4}).
The $\Omega$ possesses a richer gauge freedom than the original
$\tau$-group.
Its transformation law reads
\be
e^{{\rm i}\Omega'}=e^{{\rm i}\lambda}e^{{\rm i}\Omega}e^{-{\rm i}\tau}
\qquad \lambda=\lambda^a(\zeta,u)T^a
\label{12}
\ee
where the unconstrained analytic gauge parameters
$\lambda^a(\zeta,u)$
are real with respect to the analyticity-preserving conjugation,
${\stackrel{\smile}{\lambda}}{}^a =\lambda^a$.
The $\l$-frame is defined by
\be
\cD_{\underline{M}} \longrightarrow
\nabla_{\underline{M}}=e^{{\rm i}\Omega}{\cal D}_{\underline{M}}
e^{-{\rm i}\Omega} \qquad
\J \longrightarrow \J_\l = e^{{\rm i}\Omega} \J\;.
\label{13}
\ee
The transformation laws of the gauge covariant derivatives and matter
superfields read
\be
\nabla_{\underline{M}}' = e^{{\rm i}\lambda} \nabla_{\underline{M}}
e^{-{\rm i}\lambda} \qquad
\J_\l' = e^{{\rm i}\lambda} \J_\l \;.
\label{14}
\ee
In the $\l$-frame we have
\bea
& \nabla^+_\alpha=D^+_\alpha \qquad{\bar\nabla}^+_{\dot\alpha}=
{\bar D}^+_{\dot\alpha}
\qquad \nabla^0=D^0 \non \\
& \nabla^{\pm\pm}=e^{{\rm i}\Omega}D^{\pm\pm}e^{-{\rm i}\Omega}
=D^{\pm\pm}+{\rm i}V^{\pm\pm}
\label{15}
\eea
and the covariantly analytic superfield (\ref{11}) turns into
$\F^{(q)}_\l = \f^{(q)}$.
The connection $V^{++}=V^{++a}T^a$ proves to be
a real analytic superfield,
${\stackrel{\smile}{V}}{}^{++a}=V^{++a}$,
$D^+_\alpha V^{++}={\bar D}^+_{\dot\alpha}V^{++}=0$.
This superfield turns out to be the single unconstrained
prepotential of the
pure $N=2$ SYM theory and all other objects are expressed
in terms of it. In particular, the action of the theory reads \cite{zup}
\be
S_{{\rm SYM}}=\frac{1}{g^2} {\rm tr}\,
\int d^{12}z\sum\limits_{n=2}^\infty\frac{(-{\rm i})^n}
{n}\int du_1 \cdots du_n\frac{V^{++}(z,u_1)
\cdots V^{++}(z,u_n)}
{(u^+_1u^+_2)(u^+_2u^+_3)\cdots(u^+_nu^+_1)}\;.
\label{16}
\ee
The rules of integration over $SU(2)$, as well as the properties
of harmonic distributions, are given in refs.
\cite{gikos,gios}.

In general, the gauge superfield is coupled to $N=2$ matter
multiplets. They are described by the $q$-hypermultiplet $q^+(\zeta,u),
{\stackrel{\smile}{q}}{}^+(\zeta,u)$ and the $\omega$-hypermultiplet
$\omega (\zeta,u)$ \cite{gikos}, which are unconstrained analytic superfields
and transform in complex $R_q$ and real $R_\o$ representations of the
gauge group respectively. The massless hypermultiplet action is given by
\be
S_{{\rm MAT}}=
- \int  du \,d\zeta^{(-4)}\,
{\stackrel{\smile}{q}}{}^+ \nabla^{++}q^+ -
\frac{1}{2}\int du \,d\zeta^{(-4)}
\,\nabla^{++}\omega^{\rm T}\nabla^{++}\omega
\label{17}
\ee
where the integration is over the analytic subspace (\ref{3}).
The case when some hypermultiplets are massive corresponds to
switching on an extra coupling to a covariantly constant $N=2$
super Yang-Mills background \cite{bbiko,revis,higgs}.
The hypermultiplet mass terms can also be obtained via the
Scherk-Schwarz dimensional reduction from six dimensions
\cite{gio,zu,ikz}.

In the framework of the background field method, one splits the gauge
superfield $V^{++}$ into background $V^{++}$ and quantum $v^{++}$ parts
\be
V^{++}\rightarrow V^{++}+g\,v^{++}
\label{18}
\ee
The theory (\ref{16}) is quantized by imposing background covariant
gauge conditions in order to obtain a gauge invariant effective
action. This procedure has been carried out in \cite{back}.
The theory possesses two types of unconstrained analytic ghosts;
the anticommuting Faddeev-Popov ghosts ${\bf b}(\z,u)$ , ${\bf c}(\z,u)$
and the commuting Nielsen-Kallosh ghost $\f(\z,u)$, all in the adjoint
representation of the gauge group. The quantum action reads
\be
S_{{\rm QUANT}} = S_{2} + S_{{\rm INT}}
\label{19}
\ee
where
\bea
S_{2} &=&
-\frac{1}{2} {\rm tr}\,\int du \,d\zeta^{(-4)}\,
v^{++}{\stackrel{\frown}{\Box}}{}_\l v^{++}+{\rm tr}\,\int du \,
d\zeta^{(-4)}\,
{\bf b}(\nabla^{++})^2{\bf c} \non \\
&{}& + \frac{1}{2}{\rm tr}\,
\int du \,d\zeta^{(-4)}\,\phi(\nabla^{++})^2\phi
\label{20} \\
S_{{\rm INT}} &=&
-{\rm tr}\,\int d^{12}z
\sum\limits_{n=3}^\infty  {\frac{(-{\rm i}g)}{n}}^{n-2} \int du_1\cdots du_n
\frac{v^{++}_\tau(z,u_1)\cdots
v^{++}_\tau(z,u_n)}{(u^+_1u^+_2)(u^+_2u^+_3)\cdots(u^+_nu^+_1)} \non\\
&{}&  -{\rm i}\,g\, {\rm tr} \,\int
du \,d\zeta^{(-4)}\, \nabla^{++}{\bf b}\;[v^{++}, {\bf c}]\;.
\label{21}
\eea
Here $v^{++}_\tau$ denotes the background $\t$-transform of $v^{++}$
\be
v_\tau^{++}=e^{-{\rm i}\Omega}v^{++}e^{{\rm i}\Omega}
\label{22}
\ee
and ${\stackrel{\frown}{\Box}}{}_\l$ the $\l$-transform of the
analytic d'Alembertian\footnote{
We use the notation
$(\cD^+)^4 = \frac{1}{16} (\cD^+)^2 ({\bar \cD}^+)^2$,
$(\cD^\pm)^2=\cD^{\pm \alpha} \cD^\pm_\alpha$,
$({\bar \cD}^\pm)^2 = {\bar \cD}^\pm_{\dot{\alpha}}
{\bar \cD}^{\pm \dot{\alpha}}$
and similar notation for the flat derivatives.}
\be
{\stackrel{\frown}{\Box}}{} = -\frac{1}{2}(\cD^+)^4(D^{--})^2
\label{23}
\ee
which takes the second-order form
\bea
{\stackrel{\frown}{\Box}}{}&=&
{\cal D}^m{\cal D}_m+
\frac{{\rm i}}{2}({\cal D}^{+\alpha}W){\cal D}^-_\alpha+\frac{{\rm i}}{2}
({\bar{\cal D}}^+_{\dot\alpha}{\bar W}){\bar{\cal D}}^{-{\dot\alpha}}-
\frac{{\rm i}}{4}({\cal D}^{+\a} {\cal D}^+_\a W) D^{--}\non \\
&{}& +\frac{{\rm i}}{8}[{\cal D}^{+\alpha},{\cal D}^-_\alpha] W
+ \frac{1}{2}\{{\bar W},W \}
\label{24}
\eea
when acting on the covariantly analytic superfields.
Using the Bianchi identities
\be
\cD^{+\a} \cD^+_\a W = {\bar \cD}^+_\ad {\bar \cD}^{+\ad} \bar W \qquad
[ \cD^{+\a}, \cD^-_\a ] W = [{\bar \cD}^+_\ad, {\bar \cD}^{-\ad}] \bar W
\label{25}
\ee
one can present ${\stackrel{\frown}{\Box}}$ in a slightly different form.
The effective action is defined by the path integral representation
\cite{back}
\be
e^{{\rm i}\Gamma_{{\rm SYM}}}=e^{{\rm i}S_{{\rm SYM}}}
\int{\cal D}v^{++}{\cal D}{\bf b}
{\cal D}{\bf c}{\cal D}\phi\,
\left({\rm Det}_{(4,0)} \,
{\stackrel{\frown}{\Box}}{}_\l \right)^\hf \,
e^{{\rm i}S_{{\rm QUANT}}}
\label{ef}
\ee
where ${\rm Det}_{(4,0)}\, {\stackrel{\frown}{\Box}}{}_\l$ corresponds
to the following functional integral over anticommuting analytic
superfields $\r^{(4)}(\z,u)$ and $\s(\z,u)$
\be
{\rm Det}_{(4,0)} \,
{\stackrel{\frown}{\Box}}{}_\l =
\int {\cal D}\r^{(4)} {\cal D}\s
\exp \;\left\{ {\rm i} \;{\rm tr} \int du \, d\zeta^{(-4)} \,
\r^{(4)} {\stackrel{\frown}{\Box}}{}_\l \,\s \right\} \;.
\label{gav}
\ee

The background-quantum splitting (\ref{18}) should be accompanied by
similar splitting for the matter superfields. Within the background
field method, the effective action is described by the vacuum diagrams only,
and the propagators and vertices are background dependent. The ghost
superfields originate in the internal lines. In accordance with (\ref{20}),
the Nielsen-Kallosh ghost contributes to the one-loop effective action only.
For the general $N=2$ SYM theory with classical action
$S_{{\rm SYM}} + S_{{\rm MAT}}$,
our strategy will consist of
inserting all terms from $S_{{\rm QUANT}}$
with the matter background superfields
into $S_{{\rm INT}}$.

The one-loop correction
should be investigated separately, since it is given in terms of functional
determinants of special differential operators.
The purely Yang-Mills part $\G^{(1)}[V^{++}]$
of the one-loop effective action $\G^{(1)}$ is given by
\bea
\G^{(1)}[V^{++}] &=& S_{{\rm SYM}} \non \\
& & + {\rm i}\,{\rm Tr}\,{}_{R_q}\, \ln (\nabla^{++}) +
\frac{{\rm i}}{2}\,{\rm Tr}\,{}_{R_\o}\, \ln (\nabla^{++})^2
-\frac{{\rm i}}{2}\,{\rm Tr}\,{}_{ad}\, \ln
(\nabla^{++})^2 \non \\
& & + \frac{{\rm i}}{2}\,{\rm Tr}\,{}_{(2,2)} \, \ln
{\stackrel{\frown}{\Box}}{}_\l
- \frac{{\rm i}}{2}\,{\rm Tr}\,{}_{(4,0)} \, \ln
{\stackrel{\frown}{\Box}}{}_\l \;.
\label{26}
\eea
Here the second line includes  the contributions
from the matter hypermultiplets and the ghost superfields,
respectively. The first term in the third line comes from the
functional integral
\be
\left( {\rm Det}_{(2,2)} \,
{\stackrel{\frown}{\Box}}{}_\l \right)^{-\hf} =
\int {\cal D}v^{++} \,
\exp \, \left\{ -\frac{{\rm i}}{2}\, {\rm tr}\, \int du \, d\zeta^{(-4)} \,
v^{++} {\stackrel{\frown}{\Box}}{}_\l \,v^{++} \right\} \;.
\ee
One possible prescription for calculating the functional determinants
in the second line of (\ref{26}) has been given
in our paper \cite{bbiko}.
These one-loop contributions to the effective action contain
all information about the ultraviolet divergences
of the general $N=2$ SYM theory, since
the one-loop supergraphs with matter external lines, as well as all the
higher loop supergraphs, will be shown to be
ultravioletly finite. The functional determinants in the third line of
(\ref{26}) can produce only ultravioletly finite corrections to the
effective action. Therefore we are not going to discuss here the
one-loop effective action and concentrate our attention only on
higher-loop corrections to effective action.

${}$From eqs. (\ref{17}) and (\ref{20}), one can derive the superfield
propagators in the $\l$-frame
(all indices are suppressed)
\bea
<v^{++}(1)\,v^{++}(2)>
&=& - \frac{{\rm i}}{{\stackrel{\frown}{\Box}}{}}_\l
{\stackrel{\longrightarrow}{(D_1^+)^4}}
\biggl\{ \delta^{12}(z_1-z_2) \d^{(-2,2)}(u_1,u_2) \biggr\} \non\\
&=& - \frac{{\rm i}}{{\stackrel{\frown}{\Box}}{}}_\l
\biggl\{ \delta^{12}(z_1-z_2) \d^{(-2,2)}(u_1,u_2) \biggr\}
{\stackrel{\longleftarrow}{(D_1^+)^4}}
\non \\
<q^+(1)\,{\stackrel{\smile}{q}}{}^+(2)>
&=& \; \frac{{\rm i}}{{\stackrel{\frown}{\Box}}{}}_\l
{\stackrel{\longrightarrow}{(D_1^+)^4}}{}
\left\{ e^{{\rm i}\Omega (1)}\delta^{12}(z_1-z_2)
{1\over (u^+_1 u^+_2)^3}e^{-{\rm i}\Omega (2)} \right\}
{\stackrel{\longleftarrow}{(D_2^+)^4}}
\non \\
<\omega(1)\,\,\omega^{\rm T}(2)>
&=& - \frac{{\rm i}}{{\stackrel{\frown}{\Box}}{}}_\l
{\stackrel{\longrightarrow}{(D_1^+)^4}}{}
\left\{ e^{{\rm i}\Omega (1)}\delta^{12}(z_1-z_2)
{(u^-_1 u^-_2)\over (u^+_1 u^+_2)^3}
e^{-{\rm i}\Omega (2)}\right\}
{\stackrel{\longleftarrow}{(D_2^+)^4}}
\non \\
<{\bf c}(1)\,\,{\bf b}(2)>
&=& - \frac{{\rm i}}{{\stackrel{\frown}{\Box}}{}}_\l
{\stackrel{\longrightarrow}{(D_1^+)^4}}{}
\left\{ e^{{\rm i}\Omega (1)}\delta^{12}(z_1-z_2)
{(u^-_1 u^-_2)\over (u^+_1 u^+_2)^3}
e^{-{\rm i}\Omega (2)}\right\}
{\stackrel{\longleftarrow}{(D_2^+)^4}}\;.
\label{27}
\eea
Here the propagators involve the background bridge $\O$, which is a non-local
function of the gauge superfield $V^{++}$. Their structure becomes much
simpler in the $\t$-frame
\bea
<v^{++}_\t(1)\,v^{++}_\t(2)>
&=& - \frac{{\rm i}}{{\stackrel{\frown}{\Box}}{}}
{\stackrel{\longrightarrow}{(\cD_1^+)^4}}{}
\biggl\{ \delta^{12}(z_1-z_2) \d^{(-2,2)}(u_1,u_2) \biggr\}
\non \\
<q^+_\t(1)\,{\stackrel{\smile}{q}}{}^+_\t(2)>
&=& \; \frac{{\rm i}}{{\stackrel{\frown}{\Box}}{}}
{\stackrel{\longrightarrow}{(\cD_1^+)^4}}{}
\left\{ \delta^{12}(z_1-z_2)
{1\over (u^+_1 u^+_2)^3} \right\}
{\stackrel{\longleftarrow}{(\cD_2^+)^4}}
\non \\
<\omega_\t(1)\,\,\omega^{\rm T}_\t(2)>
&=& - \frac{{\rm i}}{{\stackrel{\frown}{\Box}}{}}
{\stackrel{\longrightarrow}{(\cD_1^+)^4}}{}
\left\{ \delta^{12}(z_1-z_2)
{(u^-_1 u^-_2)\over (u^+_1 u^+_2)^3}
\right\}
{\stackrel{\longleftarrow}{(\cD_2^+)^4}}
\non \\
<{\bf c}_\t(1)\,\,{\bf b}_\t(2)>
&=& - \frac{{\rm i}}{{\stackrel{\frown}{\Box}}{}}
{\stackrel{\longrightarrow}{(\cD_1^+)^4}}{}
\left\{ \delta^{12}(z_1-z_2)
{(u^-_1 u^-_2)\over (u^+_1 u^+_2)^3}
\right\}
{\stackrel{\longleftarrow}{(\cD_2^+)^4}}\;.
\label{28}
\eea
It is seen that in the $\t$-frame, the propagators depend on the gauge
superfield $V^{++}$ {\it only} via the $u$-independent connection $A_M$
specifying the gauge-covariant derivatives (\ref{5}). This
property of the propagators in the $\tau$-frame turns out to be very useful
for the investigation of the divergence structure.

We now present the proof of the $N=2$ non-renormalization theorem.
Consider the loop expansion of the effective action within the
context of the background
field method. As is well known, the effective action in this framework
is given by vacuum diagrams (that is, diagrams without external
lines) with background field dependent propagators and vertices (see,
for example \cite{bos}). In our case, the corresponding propagators are defined
by eqs. (\ref{27}) and (\ref{28}), and the vertices can be read off from eqs.
(\ref{21}) and (\ref{17}). It is evident that any such
diagram can be expanded in terms of background fields, and leads to a set of
conventional diagrams with an arbitrary number of external
legs. To obtain the propagators and vertices for these conventional
diagrams, we should switch off the background fields in eqs. (\ref{27}),
(\ref{28}), (\ref{21}) and  (\ref{17}). As a result,
we arrive at conventional harmonic
supergraphs, the fundamentals of which were formulated in ref.
\cite {gios}. The third ghost $\phi$ completely decouples. We now discuss
some useful features of the above supergraphs.

As follows from eqs. (\ref{17}) and (\ref{21}), the gauge superfield vertices
are given by integrals over the full superspace, while the matter
vertices and the Faddeev-Popov ghosts vertices are given by integrals
over the analytic subspace. Note, however, that propagators (\ref{27}) and
(\ref{28}) contain factors of $(D^+)^4$, which can be used to
transform integrals over the analytic subspace into integrals over
the full superspace if we make use of the identity
\be \int du \,
d\z^{(-4)}\, (D^+)^4 \cL =  \int d^{12}z \,du \, \cL \label{29}
\ee
The cost of doing this is, as a rule, the removal of one of
the two $(D^+)^4$-factors entering each matter and ghost propagator
(\ref{27}).  There is, however, one special case. Let us consider a vertex
with two external $\o$-legs, and start to transform the corresponding
integral over the analytic subspace  into an integral over the full
superspace. To do this, we should remove the factor $(D^+)^4$
from one of the two gauge superfield propagators (\ref{27}) associated
with this vertex.  As a result of transforming all integrals
over the analytic subspace into integrals over the full superspace,
each of the remaining propagators will contain, at most, one
factor of $(D^+)^4$. Some applications of this procedure to the
calculation of concrete harmonic supergraphs were considered in refs.
\cite{gios,bbiko}. Thus, any supergraph contributing to the effective action
is given in terms of the integrals over the full $N=2$ harmonic superspace. Since
this conclusion is true for each conventional supergraph in the
expansion of a given background field supergraph,
we see that an arbitrary background
field supergraph is also given by integrals over the full $N=2$
harmonic superspace. This is in complete analogy with $N=1$
supersymmetric field theories, where an arbitrary supergraph contributing
to the effective action in the background field method contains only
integrals over the full $N=1$ superspace, but not over the chiral subspace(see,
for example \cite{grs,ggrs,west,bk}) .

Once we have constructed the supergraphs with all vertices integrated
over the full $N=2$ harmonic superspace, we can perform all but one of
the integrals over the $\theta$'s, step by step and loop by loop, due to
the spinor delta-functions $\d^8(\q_i-\q_j)$ contained in the
propagators (\ref{27}). To do this, we
remove the $(D^+)^4$-factors acting on the spinor delta-functions in the
propagators by making an integration by parts. This allows one to obtain spinor
delta-functions without $(D^+)^4$-factors. One can then perform the
integrals over the $\q$'s. We
note that in the process of integration by parts, some of the
$(D^+)^4$-factors can act on the external legs of the supergraph.
To obtain a non-zero result in the case of
an $L$-loop supergraph, we should remove $2L$ factors of $(D^+)^4$
attached to some of the propagators using the identity \cite{gios}
\be
\d^8(\q_1 - \q_2) (D^+_1)^4(D^+_2)^4\, \d^8(\q_1 - \q_2)
=(u^+_1u^+_2)^4 \d^8(\q_1 - \q_2) \;.
\label{30}
\ee  
(explicit examples of this procedure can be found in
refs. \cite{gios,bbiko}). Thus, any supergraph
contributing to the effective action is given by a single integral over
$d^8\q$. We see again the complete analogy, at each step, with $N=1$
supersymmetric field theories (see, for example \cite{grs,ggrs,west,bk}).

The next step in our investigation is the calculation of the superficial
degree of divergence for the theory under consideration. Let us consider an
$L$-loop supergraph $G$ with $P$ propagators, $N_{MAT}$ external matter
legs and an arbitrary number of gauge superfield external legs. We
denote by $N_D$ the number of spinor covariant derivatives acting on
the external legs as a result of integration by parts in the process of
transformating the contributions to a single integral over $d^8\q$. The
superficial degree of divergence $\o(G)$ of the supergraph $G$ can
readily be found
\bea
\o(G) &=& 4L - 2P +(2P - N_{\rm MAT} -4L) - \hf N_D \non \\
&=& - N_{\rm MAT} - \hf N_D \;.
\label{31}
\eea  
Here $4L$ is the contribution of the integrals over momenta, $-2P$
comes from the factors $\Box^{-1}$ contained in the propagators and
$2P-N_{MAT}$ is the contribution of the factors $(D^+)^4$ associated
with the propagators. We should note that, at least, one of the two
$(D^+)^4$-factors in each matter and ghost superfield propagator
(\ref{27}) was used to restore the full $N=2$ harmonic superspace
measure $d^{12}zdu$. It follows that each of the propagators (\ref{27})
effectively has, at most, one factor of $(D^+)^4$, leading to the contribution
$2P$ in eq. (\ref{31}). The contribution $-4L$ arises from the fact
that the factors
$(D_1^+)^4(D_2^+)^4$ in the propagators were removed
using equation eqs. (\ref{30}) in
each of the $L$ loops. However, if the supergraph has external matter legs,
the actual number of $(D^+)^4$  factors in the propagator will be less
then we counted above. Let us start with two examples given in Fig. 1 and
Fig. 2

\vspace{3mm}
\mov(1.5,0.0){
\vDot\lin(-.7,0.7)\lin(-0.7,-0.7)\ind(-8.0,7.0){\omega}
\ind(-9.0,-7.0){\omega}\mov(-0.1,0.0){\Circle(0.3)}
\ind(8.5,5.0){|}\ind(10.0,8.0){(D^+)^4}
\mov(-0.3,0.0){\wavearcto(2.1,0.0)[+]}
\ind(7.0,-5.0){|}\ind(7.0,-9.0){(D^+)^4}
\mov(-0.5,0.0){\wavearcto(2.1,0.0)[-]}
\mov(1.5,0.0){\vDot\lin(0.7,0.7)\lin(0.7,-0.7)\ind(8.5,7.0){\omega}
\ind(7.5,-7.0){\omega}\mov(-0.1,0.0){\Circle(0.3)}}
\ind(28.0,0.0){\Longrightarrow}
\mov(4.0,0.0){
\vDot\lin(-.7,0.7)\lin(-0.7,-0.7)\ind(-8.0,7.0){\omega}
\ind(-9.0,-7.0){\omega}
\mov(-0.2,0.0){\wavearcto(1.9,0.0)[+]}
\mov(-0.3,0.0){\wavearcto(1.9,0.0)[-]}
\mov(1.5,0.0){\vDot\lin(0.7,0.7)\lin(0.7,-0.7)\ind(8.5,7.0){\omega}
\ind(7.5,-7.0){\omega}}}
}

\vspace{3mm}
{}\hspace{6.5cm} Fig. 1

\mov(1.5,0.0){
\vDot\lin(-.7,0.0)\ind(-8.0,0.0){q}
\mov(-0.1,0.0){\Circle(0.3)}
\ind(8.5,5.0){|}\ind(10.0,8.0){(D^+)^4}
\mov(-0.3,0.0){\wavearcto(2.1,0.0)[+]}
\ind(10.5,-5.0){|}\ind(11.0,-9.0){(D^+)^4}
\ind(2.5,-5.0){|}\ind(3.0,-9.0){(D^+)^4}
\mov(-0.5,0.0){\arcto(2.1,0.0)[-]}
\mov(1.5,0.0){\vDot\lin(0.7,0.0)\ind(8.5,0.0){q}
\mov(-0.1,0.0){\Circle(0.3)}}
\ind(28.0,0.0){\Longrightarrow}
\mov(4.0,0.0){
\vDot\lin(-.7,0.0)\ind(-8.0,0.0){q}
\ind(8.5,4.5){|}\ind(10.0,7.5){(D^+)^4}
\mov(-0.1,0.0){\wavearcto(1.8,0.0)[+]}
\mov(-0.2,0.0){\arcto(1.8,0.0)[-]}
\mov(1.5,0.0){\vDot\lin(0.7,0.0)\ind(8.5,0.0){q}
}}
}

\vspace{3mm}
{}\hspace{6.5cm} Fig. 2

\noindent Here \put(16,3){\circle{10.00}}\put(13,0){$\bullet$}
\hspace{1cm} means a vertex corresponding to the integral over the analytic
subspace and $\bullet$ means the same vertex transformed into an
integral over the full superspace. The solid line corresponds
to a matter superfield propagator. Fig. 1 shows that, in the process of
the transformation, we removed all $(D^+)^4$-factors from the gauge
propagators. Fig. 2 shows that, in process of transformation, we removed
two factors of $(D^+)^4$ from the matter propagator. These examples
illustrate the general situation that each two external matter legs take
away one $(D^+)^4$-factor from the integrand.  Indeed, let us consider
a chain of propagators which ends at two external $q^+$- or $\o$-legs.
Taking into account that any interaction in the theory under
consideration necessarily includes gauge superfields, one observes
that each of the above chains contains a number of vertices which is larger
than the number of matter propagators by one.  As a result, after
restoring the full measure, we get the number of remaining
$(D^+)^4$-factors to be equal to the number of propagators minus one. This
means that the two external matter legs take away one factor of $(D^+)^4$ from
the integrand. This result explains the term $-N_{MAT}$
in eq. (\ref{31}). In the process of integration by parts in order
to restore the full measure, some of the spinor derivatives can act on the
external legs. Hence, they can not influence the  power of momentum in the
integrand. This leads to the contribution $-\frac{1}{2}N_D$ in eq. (\ref{31}).
We see immediately that all supergraphs with external matter legs are automatically
finite. As to supergraphs with pure gauge superfield legs, they are
clearly finite only if some non-zero number of spinor covariant
derivatives acts on the external legs. 
We will now show that this is always the case
beyond one loop.

The Feynman rules for $N=2$ supersymmetric field theories in the harmonic
superspace approach have been formulated in the $\l$-frame,
where the propagators are given by (\ref{27}). As we have noted, all
vertices in the background field supergraphs, including the vertices
of matter and Faddeev-Popov ghosts superfields, can be given in a form
containing integrals over the full $N=2$ harmonic superspace only. To be
more precise, this property is stipulated by the identity in $\l$-frame
\be
(D^+)^4 \;{\stackrel{\frown}{\Box}}{}_\l =
{\stackrel{\frown}{\Box}}{}_\l \;(D^+)^4 \;.
\label{32}
\ee 
This identity allows one to operate with
factors $(D^+)^4$ as in case without background field, and use them to
transform the integrals over the analytic subspace into integrals over
the full superspace directly in background field supergraphs. Let us
consider the structure of the propagators in the $\l$-frame (\ref{27}). 
The background field $V^{++}$ enters these propagators via both
${\stackrel{\frown}{\Box}}{}_\l$ and the background bridge $\O$. The
form of the propagators (\ref{27}) has one drawback: if we use this form, 
we can not say how many spinor derivatives act on 
the external legs since the explicit
dependence of $\O$ on the background field is rather complicated. To clarify
the
situation when a number of spinor derivatives act on external legs, we use
a completely new (in comparison with conventional harmonic supergraph
approach \cite{gios}) step and transform the supergraph
to the $\t$-frame (after restoring the full superspace measure
at the matter and ghost vertices). 
The propagators in the $\t$-frame are given
by (\ref{28}); they contain, at most, one factor of $(\cD^+)^4$
after restoring the full superspace measure 
at the matter and ghost vertices.
The essential feature of these propagators is that they
contain the background field $V^{++}$ only via the
${\stackrel{\frown}{\Box}}$ and $\cD^+$-factors; that is, only via the
$u$-independent connections $A_M$ (\ref{5}) (see eqs. (\ref{23}),
(\ref{24})). But all connections $A_M$ contain at least one spinor
covariant derivative acting on the background superfield $V^{++}$
\cite{gikos}. Therefore, if we expand any background field
supergraph in the background superfield $V^{++}$, we see that each
external leg must contain at least one spinor covariant derivative.
Thus, the number $N_D$ in eq. (\ref{31}) must be greater than or equal to one.
As a conseqence $\o(G)<0$ and, hence, all supergraphs are ultravioletly
finite beyond the one-loop level. This completes the proof of the
non-renormalization theorem.

The background field formulation allows us to prove some important
properties of the quantum corrections to those parts of
the effective action which depend on the pure $N=2$ Yang-Mills
superfield $V^{++}$. As in conventional quantum field theory, we can
suppose that $\G[V^{++}]$ is described in terms of effective
Lagrangians. That is
\be
\G[V^{++}] = \left(\int d^4 x d^4\theta \, {\cL}^{(c)}_{eff}+
{\rm c.c.} \right)+\int d^4 x d^8\theta \, {\cL}_{eff}
\label{33}
\ee
where ${\cL}^{(c)}_{eff}$ can be called the chiral effective Lagrangian
and ${\cL}_{eff}$ can be called the general effective Lagrangian. If
the theory under consideration is formulated within the background
field method, the effective Lagrangians ${\cL}^{(c)}_{eff}$ and
${\cL}_{eff}$ should be constructed only from field strengths $W$ and
$\bar{W}$ and their covariant derivatives. Therefore, the effective
Lagrangians can be written as follows: ${\cL}_{eff}=H(W,\bar{W})+$
terms depending on covariant derivatives of $W$ and $\bar{W}$ and
${\cL}^{(c)}_{eff}=F(W)+$ terms depending on covariant derivatives of
the strengths and preserving chirality, with holomorphic $F(W)$ and
hermitian $H(W,\bar{W})$ functions of the superfield strengths. The
chiral effective Lagrangian of the form ${\cL}^{(c)}_{eff}=F(W)$ is
associated with the leading low-energy behaviour of the effective
action and defines the vacuum structure of the theory
\cite{hol,seiberg,sw,n1cal}. We note that the effective holomorphic
Lagrangian ${\cal L}^{(c)}$ is analogous to the chiral effective
Lagrangian in $N=1$ theories \cite{chiral}. The general effective
Lagrangian of the form ${\cL}_{eff}=H(W,\bar{W})$ defines the first
non-leading corrections to the effective dynamics
\cite{hen,n1cal,bbiko,ds}.

A simple consequence of the background field formulation is that there
are no quantum corrections to $H(W,\bar{W})$ at two loops in the pure $N=2$
super Yang-Mills theory without matter. All two-loop supergraphs
contributing to the effective action within the background field method are
given in Fig. 3

\vspace{3mm}
\mov(2.5,0){\wavecirc(1.0)\mov(1.0,0){\wavecirc(1.0)}
\mov(3.0,0){\wavecirc(1.0)}\mov(2.5,0){\wavelin(1.0,0)}
\mov(5.0,0){\dashcirc(1.0)}\mov(4.5,0){\wavelin(1.0,0)}
}

{}\hspace{7cm} Fig. 3

\noindent Here the wavy line corresponds to the super Yang-Mills propagator
and the dotted line to the ghost
propagator. These propagators are given by eqs. (\ref{27}) and (\ref{28}).

As we have noted, in order to get a non-zero result in two-loop
supergraphs, we should use eq. (\ref{30}) twice. This implies that 
we should have
16 spinor covariant derivatives to reduce the $\q$-integrals over the full
superspace to a single one. All these spinor derivatives come from the
propagators (\ref{27}) and (\ref{28}). After we 
use one $(D^+)^4$-factor from
the ghost propagator to restore the full superspace measure, 
we see that the
propagators of both gauge and ghost superfields have at most 
a single factor
$(D^+)^4$. It is evident that the number of these $D$-factors is
not sufficient to form all 16 $D$-factors we need in two-loop supergraphs.
However, there is another source of $D$-factors in supergraphs. Extra
$D$-factors can come from the expansion of 
the inverse analytic d'Alembertian
(\ref{24}) in a power series of the field-strengths $W$ and
$\bar{W}$. As can be seen
from (\ref{24}), the spinor covariant derivatives enter the analytic
d'Alembertian always multiplied by the derivatives of $W$ and
$\bar{W}$. If we omit these derivatives, the operator
$\stackrel{\frown}{\Box}$ in (\ref{24}) takes the form
$\stackrel{\frown}{\Box}={\cD^m}{\cD_m}+\frac{1}{2}\{\bar{W},W\}$, and
does not contain the spinor covariant derivatives. Therefore, the
two-loop supergraphs given in Fig. 3 do not contribute to the effective
action if the covariant derivatives of $W$ and $\bar{W}$ are switched off.
Thus, there are no two-loop quantum corrections to the
non-holomorphic effective Lagrangian $H(W,\bar{W})$. It is worth
pointing out that this result is simply a consequence of the $N=2$
background field method and does not demand any direct calculation of
the supergraphs.  Moreover, this result will be true even if we take
into account the two-loop matter contribution to the effective action
depending only on $V^{++}$. This is almost obvious since, after
restoring the full superspace measure, the matter superfield
propagators have effectively the same structure as the gauge and ghost
superfield propagators. Another consequence of the $N=2$ background
field method is a very simple proof of the known result concerning the
absence of corrections to $F(W)$ beyond one loop. We will consider this
last statement in a forthcoming paper.

To conclude, we have presented a rigorous and simple proof of the $N=2$
non-renormalization theorem according to which the divergences in
$N=2$ super Yang-Mills theory with matter are absent beyond one loop.
Our proof was based on two key
details.  The first is the formulation of the theory in harmonic superspace
in terms of unconstrained superfields. As a result, we have no
quantization problems, as compared to the formulations in conventional
$N=2$ superspace. The Feynman rules have a simple structure
analogous to those in $N=1$ supersymmetric theories. Second, the
background field method \cite{back} allows one to formulate a manifestly $N=2$
supersymmetric and gauge invariant perturbation procedure for
calculating the effective action. The most important point of our
proof was the transformation to the $\t$-frame, where the entire dependence
of the propagators on the background gauge superfield was contained in the
covariant derivatives.

The background field method gives the possibility to investigate the
structure of the effective action in a very clear and simple manner. In
particular, we have shown, without the necessity of a direct
calculation, that there
are no two-loop corrections to the effective Lagrangian $H(W,\bar{W})$.

\vspace{1cm}

\noindent
{\bf Acknowledgements.}
We are grateful to E.A. Ivanov for fruitful discussions of
quantum aspects of the harmonic superspace approach.
ILB and SMK acknowledge
partial support from the RFBR-DFG project No. 96-02-001800
and the RFBR project No. 96-02-16017.
ILB also thanks the University
of Pennsylvania  Research Foundation for partial support.
BAO acknowledges the DOE Contract No. DE-AC02-76-ER-03072
for partial support. SMK and BAO are grateful to the Alexander von
Humboldt Foundation for partial support.
ILB is grateful to the Institute for Physics, Humboldt University of Berlin,
where this work was started, and to Prof. L\"ust for hospitality.
He also thanks the Institute for Theoretical Physics, University
Hannover and the Institute for Theoretical Physics, Swiss Federal
Technical Institute, Zurich for hospitality and is grateful to
Prof. Dragon for interesting discussions.


\begin{thebibliography}{99}
\bibitem{m} L. Mezincescu, JINR report P2-12572 (1979).
\bibitem{k} J. Koller, Nucl. Phys. B222 (1983) 319;
Phys. Lett. B124 (1983) 324.
\bibitem{hst} P.S. Howe, K.S. Stelle and P.K. Townsend, Nucl.Phys. B236
(1984) 125.
\bibitem{sg} W. Siegel and S.J. Gates, Nucl. Phys. B189 (1981) 295.
\bibitem{gikos} A. Galperin, E. Ivanov, S. Kalitzin, V. Ogievetsky and
E. Sokatchev, Class. Quant. Gravit. 1 (1984) 469.
\bibitem{bv} I.A. Batalin and G.A. Vilkovisky, Phys. Rev. D28 (1983) 2567
(E: D30 (1984) 508).
\bibitem{mar} S. Marculescu, Phys. Lett. B188 (1987) 203;
Fortschr. Phys. 36 (1988) 335.
\bibitem{ggrs} S.J. Gates, M.T. Grisaru, M. Ro\v{c}ek and W. Siegel,
{\it Superspace}, Benjamin-Cummings, Reading, MA, 1983.
\bibitem{west} P. West, {\it Introduction to Supersymmetry and
Supergravity}, World Scientific, Singapore, 1990.
\bibitem{beta} P. Howe, K. Stelle and P. West, Phys. Lett. 
B124 (1983) 55;\\
P. West, {\it Supersymmetry and Finiteness},
in Proceedings of the 1983 Shelter Island II Conference on 
Quantum Field Theory and Fundamental Problems of Physics,
edited by R. Jackiw, N. Kuri, S. Weinberg and E. Witten (M.I.T. Press).
\bibitem{n1cal} B. de Wit, M.T. Grisaru and M. Ro\v{c}ek, Phys.Lett. B374
(1996) 297; \\
A. Pickering and P. West, Phys.Lett. B383 (1996) 54; \\
M.T. Grisaru, M. Ro\v{c}ek and R. von Unge, Phys. Lett. B383
(1996) 415; \\
T.E. Clark and S.T. Love, Phys. Lett. B388 (1996) 577; \\
U. Lindsr\"{o}m, F. Gonzales-Rey, M. Ro\v{c}ek and R. von
Unge, Phys. Lett. B388 (1996) 581;\\
A. De Giovanni, M.T. Grisaru, M. Ro\v{c}ek,
R. von Unge and D. Zanon, Phys. Lett. B409 (1997) 251; \\
A. Yung, Nucl. Phys. B485 (1997) 38;\\
M. Matone, Phys. Rev. Lett. 78 (1997) 1412;\\
D. Bellisai, F. Fucito, M. Matone and G. Travaglini,
Phys. Rev. D56 (1997) 5218.
\bibitem{bbiko} I.L. Buchbinder, E.I. Buchbinder, E.A. Ivanov, S.M. Kuzenko
and B.A. Ovrut, {\it Effective action of the $N=2$ Maxwell multiplet
in harmonic superspace}, Preprint IASSNS-HEP-97/6,
UPR-733T, JINR E2-97-82, ITP-UH-09/97, hep-th/9703147; Phys. Lett. B,
in press.
\bibitem{ikz} E.A. Ivanov, S.V. Ketov and B.M. Zupnik,
{\it Induced hypermultiplet self-interactions in $N=2$ gauge theories},
Preprint DESY 97-094, ITP-UH-10/97, JINR E2-97-164, hep-th/9706078.
\bibitem{giosreview} A. Galperin, E. Ivanov, V. Ogievetsky and E. Sokatchev,
{\it The Methods of Harmonic Superspace}, in preparation.
\bibitem{gios} A. Galperin, E. Ivanov, V. Ogievetsky and E. Sokatchev,
Class. Quantum Grav. 2 (1985) 601; 617.
\bibitem{hsw} P.S. Howe, K.S. Stelle and P.C. West,
Class. Quantum Grav. 2 (1985) 815.
\bibitem{back} I.L. Buchbinder, E.I. Buchbinder, S.M. Kuzenko
and B.A. Ovrut, {\it The background field method for $N=2$ super
Yang-Mills theories in harmonic superspace}, Preprint IASSNS-HEP-9732T,
UPR-745T, ITP-UH-12/97, hep-th/9704214; Phys. Lett. B, in press.
\bibitem{bk} I.L. Buchbinder and S.M. Kuzenko, {\it Ideas and Methods of
Supersymmetry and Supergravity}, IOP Publ., Bristol and Philadelphia,
1995.
\bibitem{zup} B. Zupnik, Phys. Lett. B183 (1987) 175.
\bibitem{revis} I.L. Buchbinder and S.M. Kuzenko, 
Class. Quantum Grav. 14 (1997) L157.
\bibitem{higgs} N. Dragon and S.M. Kuzenko, {\it The Higgs mechanism
in $N=2$ superspace}, Preprint ITP-UH-15/97, hep-th/9705027;
Nucl. Phys. B., in press.
\bibitem{gio} A. Galperin, E. Ivanov and V. Ogievetsky, Nucl. Phys.
B282 (1987) 74.
\bibitem{zu} B.M. Zupnik, Sov. J. Nucl. Phys. 44 (1986) 512.
\bibitem{hol}
G. Sierra and P.K. Townsend, in : Supersymmetry and Supergravity 1983,
Proc. XIXth Winter School, Karpacz, ed. B. Milewski (World Scientific,
1983), p. 396; \\
B. de Wit, P.G. Lauwers, R. Philippe, S.-Q. Su and A. Van Proyen,
Phys. Lett. B134 (1984) 37;\\
S.J. Gates, Nucl. Phys. B238 (1984) 349.
\bibitem{seiberg} N. Seiberg, Phys. Lett. B206 (1988) 75.
\bibitem{sw} N. Seiberg and E. Witten, Nucl. Phys. B426 (1994) 19;
B430 (1994) 485.
\bibitem{chiral} P. West, Phys. Lett. B261 (1991) 396;\\
I.L. Buchbinder, S.M. Kuzenko and A.Yu. Petrov,
Phys. Lett. B321 (1994) 372.
\bibitem{hen} M. Henningson, Nucl. Phys. B458 (1996) 445.
\bibitem{ds} M. Dine and N. Seiberg, Phys. Lett. B409 (1997) 239.
\bibitem{bos} I.L. Buchbinder, S.D. Odintsov and I.L. Shapiro, Effective
Action in Quantum Gravity, IOP Publ., Bristol and Philadelphia, 1992.
\bibitem{grs} M. Grisaru, M. Ro\v{c}ek, W. Siegel, Nucl. Phys. B159,
(1979), 429.

\end{thebibliography}
\end{document}